\documentclass[aps, twocolumn, prl, superscriptaddress]{revtex4-2}

\usepackage{graphicx}
\usepackage{amssymb}
\usepackage{braket}
\usepackage{amsmath}
\usepackage[colorlinks=true, linkcolor=blue, citecolor=blue, urlcolor=blue]{hyperref}
\usepackage{dsfont}
\usepackage{bm} 
\usepackage{physics}
\usepackage{tabularx}
\usepackage{verbatim}
\usepackage{float}
\usepackage{romannum}
\usepackage{bbold}
\usepackage{mathtools}

\begin{document}

\title{Oligonucleotide selective detection by levitated optomechanics}
\author{Timothy Wilson}
\affiliation{School of Ocean and Earth Science, Waterfront Campus, European Way, Southampton, SO14 3ZH, United Kingdom}
\author{Owen J. L. Rackham}
\affiliation{School of Biological Sciences, University of Southampton, Southampton SO17 1BJ, United Kingdom}
\author{Hendrik Ulbricht}
\affiliation{School of Physics and Astronomy, University of Southampton, Southampton SO17 1BJ, United Kingdom}

\begin{abstract}
  This study examines the detection of oligonucleotide-specific signals in sensitive optomechanical experiments. Silica nanoparticles were functionalized using ZnCl$_2$ and 25-mers of single-stranded deoxyadenosine and deoxythymidine monophosphate which were optically trapped by a 1550 nm wavelength laser in vacuum. In the optical trap, silica nanoparticles behave as harmonic oscillators, and their oscillation frequency and amplitude can be precisely detected by optical interferometry. The data was compared across particle types, revealing differences in frequency, width and amplitude of peaks with respect to motion of the silica nanoparticles which can be explained by a theoretical model. Data obtained from this platform was analyzed by fitting Lorentzian curves to the spectra. Dimensionality reduction detected differences between the functionalized and non-functionalized silica nanoparticles. Random forest modeling provided further evidence that the fitted data were different between the groups. Transmission electron microscopy was carried out, but did not reveal any visual differences between the particle types.
\end{abstract}

\date{\today} 

\maketitle


Detecting and differentiating DNA strands has applications in the fields of medicine, data storage and evolutionary biology \cite{Campolongo2010,Doricchi2022,Nesse2010}. It is therefore of interest to develop methods to study DNA with greater speed and accuracy. The Sanger sequencing method was published in 1977 \cite{Sanger1977} with parallelization and high-throughput now standard in modern techniques \cite{Dewey2012}. This research presents an alternative method based upon the optical properties of DNA nucleotides.

Optical trapping was originally observed by Arthur Ashkin in 1970 to contain micron-sized particles \cite{Ashkin1970}. This technique, known as optical tweezing, has since found applications in biosensing and live cell imaging in a solution~\cite{Rodríguez-Sevilla2017,Tam2011}. Optical trapping in vacuum is commonly referred to as levitated optomechanics~\cite{gonzalez2021levitodynamics} and is the approach used in this work. By levitated optomechanics, it is possible to measure tiny forces of trapped particles on the order of 10$^{-20}$ N \cite{Hempston2017}, leading to the notion that a trapped particle functionalized with DNA might be distinguished from those particles that do not have surface modifications. 

Silica is routinely used in optical trapping due to its greater refraction than water at near-infrared wavelengths, a property required for stable optical trapping in a water medium \cite{Bendix2013}. Silica nanoparticles were used early in the development of optical trapping by Ashkin and Dziedzic where they demonstrated levitation of 20 $\mu$m diameter silica nanoparticles at a pressure of 1 mbar \cite{Ashkin1971,Ashkin1974,Ashkin1976}. Factors important to consider in the choice of material include high polarizability and low absorption at the wavelength of the source. Silica meets these criteria at the 1550 nm wavelength \cite{Vovrosh2018}. The properties of silica nanoparticles, and their ability to be functionalized with DNA, are the reasons why they were used in this study.

The process of DNA adsorption onto the surface of nanoparticles remains rarely studied\cite{Meng2018}. Metal ions are vital in living organisms, \and the regulation of biological processes \cite{Grazul2009} and as cofactors of DNAzymes \cite{Torabi2015}. One paper examined methods to functionalize silica nanoparticles with short fluorescein-tagged DNA strands using different metal ions as binding agents \cite{Huang2022}. They found that Zn$^{2+}$ ions from a ZnCl$_2$ solution at a concentration of 1 mM was one of the best adsorption metal ions to bind fluorescein tagged 25-mer deoxyadenosine monophosphate oligonucleotides to silica nanoparticles. This study did not carry out optical trapping of silica nanoparticles, however it did provide a foundation for the method of functionalization used in this research. 

\begin{figure}
\centering
\includegraphics[width=\linewidth]{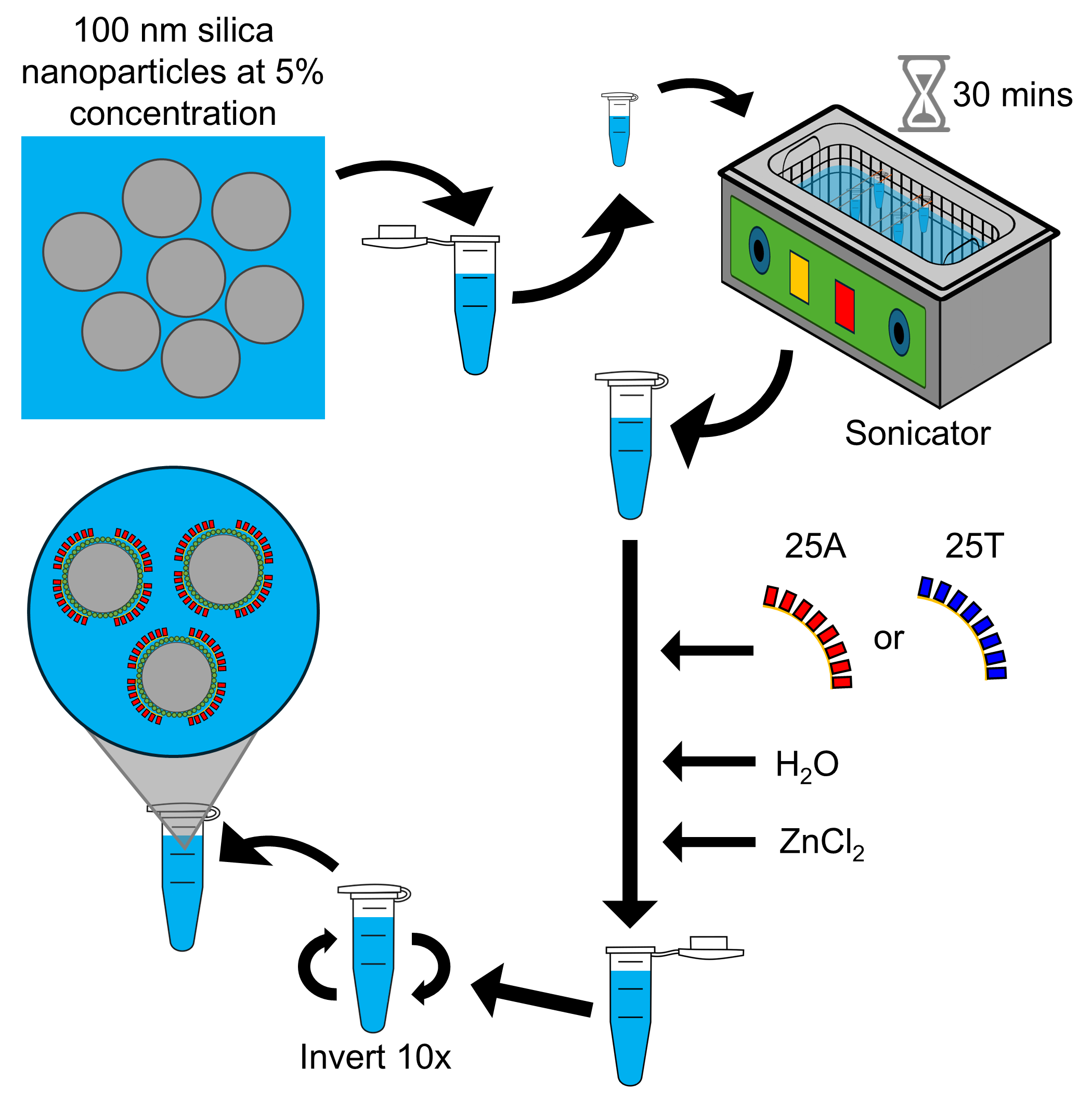}
\caption{{\bf Functionalization:} The process of functionalizing silica nanoparticles with 25-mer deoxyadenosine monophosphate (25A) or 25-mer deoxythymidine monophosphate (25T) oligonucleotides, water and ZnCl$_2$ solution. Eppendorf tube images are provided by \href{https://www.labicons.net}{Labicons}.}\label{fig1}
\end{figure}

The research presented here uses the method of particle preparation for release into the optical trap in a vacuum as shown in Figure \ref{fig1}. The sonicator is used to reduce particle aggregation when released into the optical trap from the nebulizer.

Here we present the technique of optical trapping under vacuum to explore the detection of oligonucleotide functionalization differences between groups of silica nanoparticles. Silica nanoparticles with no surface modifications (standard silica nanoparticles) were tested as a reference, then 25-mer deoxyadenosine monophosphate (25A) and 25-mer deoxythymidine monophosphate (25T) oligonucleotides were added with ZnCl$_2$ as a binding agent to the silica nanoparticles with concentrations of ZnCl$_2$ changed for the 25T variant. The raw data for each particle type was compared. Uniform Manifold Approximation and Projection (UMAP) dimensionality reduction and random forest analysis were used to examine how well these groups of particles could be classified. Transmission Electron Microscopy (TEM) imaging was also used to attempt to visualize differences. Collectively, this provides evidence for the selective detection of these silica nanoparticles using levitated optomechanics.

\begin{figure*}
    \centering
    \includegraphics[width=\linewidth]{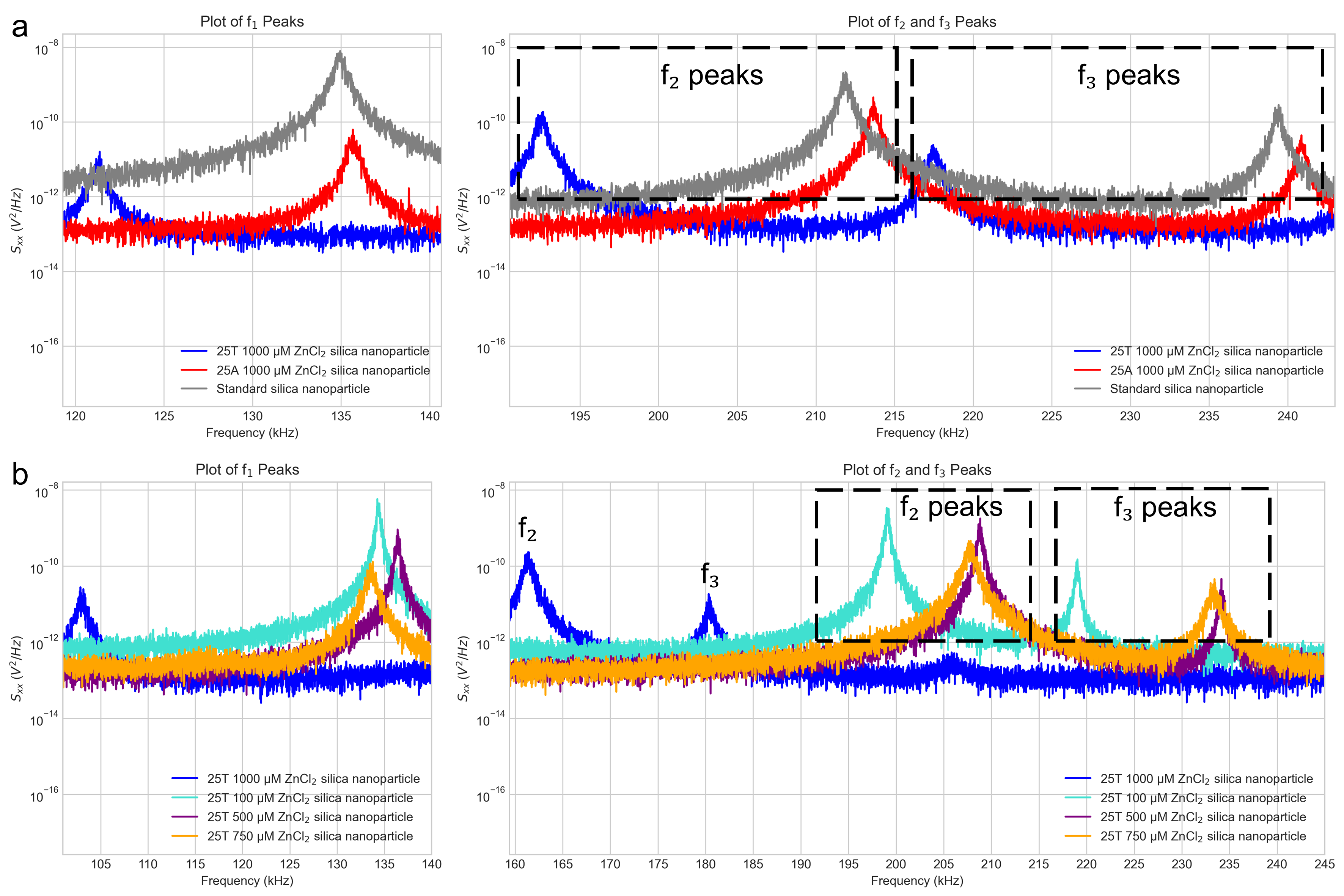}
    \caption{\textbf{PSD plots of groups of selected particle types with frequency peaks f$_1$, f$_2$ and f$_3$ displayed.} (a) Plots of the f$_1$, f$_2$ and f$_3$ peaks for one particle of each type: 25A silica nanoparticle at a 1000 $\mu$M ZnCl$_2$ concentration (red), 25T silica nanoparticle at a 1000 $\mu$M ZnCl$_2$ concentration (blue) and for a standard silica nanoparticle (grey). (b) Plots of the f$_1$, f$_2$ and f$_3$ peaks for one particle of each type: 25T silica nanoparticle at a 1000 $\mu$M ZnCl$_2$ concentration (blue), 25T silica nanoparticle at a 100 $\mu$M ZnCl$_2$ concentration (turquoise), 25T silica nanoparticle at a 500 $\mu$M ZnCl$_2$ concentration (purple) and 25T silica nanoparticle at a 750 $\mu$M ZnCl$_2$ concentration (orange).}\label{fig2}
\end{figure*}

The nanoparticles were released into a vacuum chamber, and when trapped, the chamber was pumped down to a consistent 3.5 mbar within the range of error of the pressure gauge. The Power Spectral Density (PSD) waveforms were recorded on an oscilloscope. After the data were collected, the particles were filtered to remove outliers. The process of calculating numerical columns and outlier removal was carried out by fitting Lorentzian curves to the f$_1$, f$_2$ and f$_3$ frequency peaks, which correspond to the z, x and y degrees of motion respectively, using the tool, Optoanalysis \cite{Rademacher2017}. The PSD, $S_{xx}(\omega)$, as seen in Figure \ref{fig2} has units of V$^2$/Hz and can be written as:

\begin{equation} \label{Equation 1}
S_{xx}(\omega) = \gamma^2 \frac{k_B T_0}{\pi m} \frac{\Gamma_0}{\left(\omega_0^2 - \omega^2\right)^2 + \omega^2 \Gamma_0^2}   
\end{equation}

where $\gamma$ is the conversion factor, $k_B$ is the Boltzmann constant, $T_0$ is the temperature of the environment, $m$ is mass of the particle, $\Gamma_0$ is the damping rate, $\omega_0$ is the natural angular frequency and $\omega$ is the angular frequency at which the PSD is calculated. Equation (\ref{Equation 1}) can be simplified to the experimental data, $S_{xx}^{exp}$, as below:

\begin{equation} \label{Equation 2} 
S_{xx}^{exp}=\frac{A}{{(B^2\ -\omega^2)}^2+\omega^2C^2\ }
\end{equation}

where $A=\frac{\gamma^2k_BT_0\Gamma_0}{\pi m}$, ${B=\omega}_0$ and ${C=\Gamma}_0$ are free fit parameters. The conversion factor $\gamma$ converts the PSD to units of $\frac{m^2}{Hz}$ and can be calculated as follows in equation (\ref{Equation 3}):

\begin{equation} \label{Equation 3} 
\gamma=\ \sqrt{\frac{A}{C}\frac{\pi m}{k_BT_0}}
\end{equation}

The nanoparticle is assumed to be in a thermal equilibrium where $T_0$ = 300 K. The radius is derived from the fit parameter of the Lorentzian curve from the PSD in equation (\ref{Equation 4}) as in the literature\cite{Vovrosh2018}:

\begin{equation} \label{Equation 4} 
r\ =\ 0.619\frac{9\pi}{\sqrt2}\frac{\eta_{air}d^2}{\rho k_BT_0}\frac{P_{gas}}{C}
\end{equation}

where $r$ is the particle radius, $\eta_{air}$ is the viscosity of air, $d$ is the diameter of the atmospheric particles, $P_{gas}$ is the pressure measured from the pressure sensor and $\rho$ is the particle material density. Assuming the particle to be spherical, the mass of the particle can be calculated using $m = \frac{4}{3}\pi r^3$.

The resulting parameters were filtered and outliers removed. An outlier is defined as being greater than 1.5 interquartile ranges from the median for any numerical column. The complete raw dataset is available in the supplementary data. There are considerable variances in the size of silica nanoparticles, which is the primary reason for this filtering. The data analysis presented here are with outliers removed.

\begin{figure*}
    \centering
    \includegraphics[width=\linewidth]{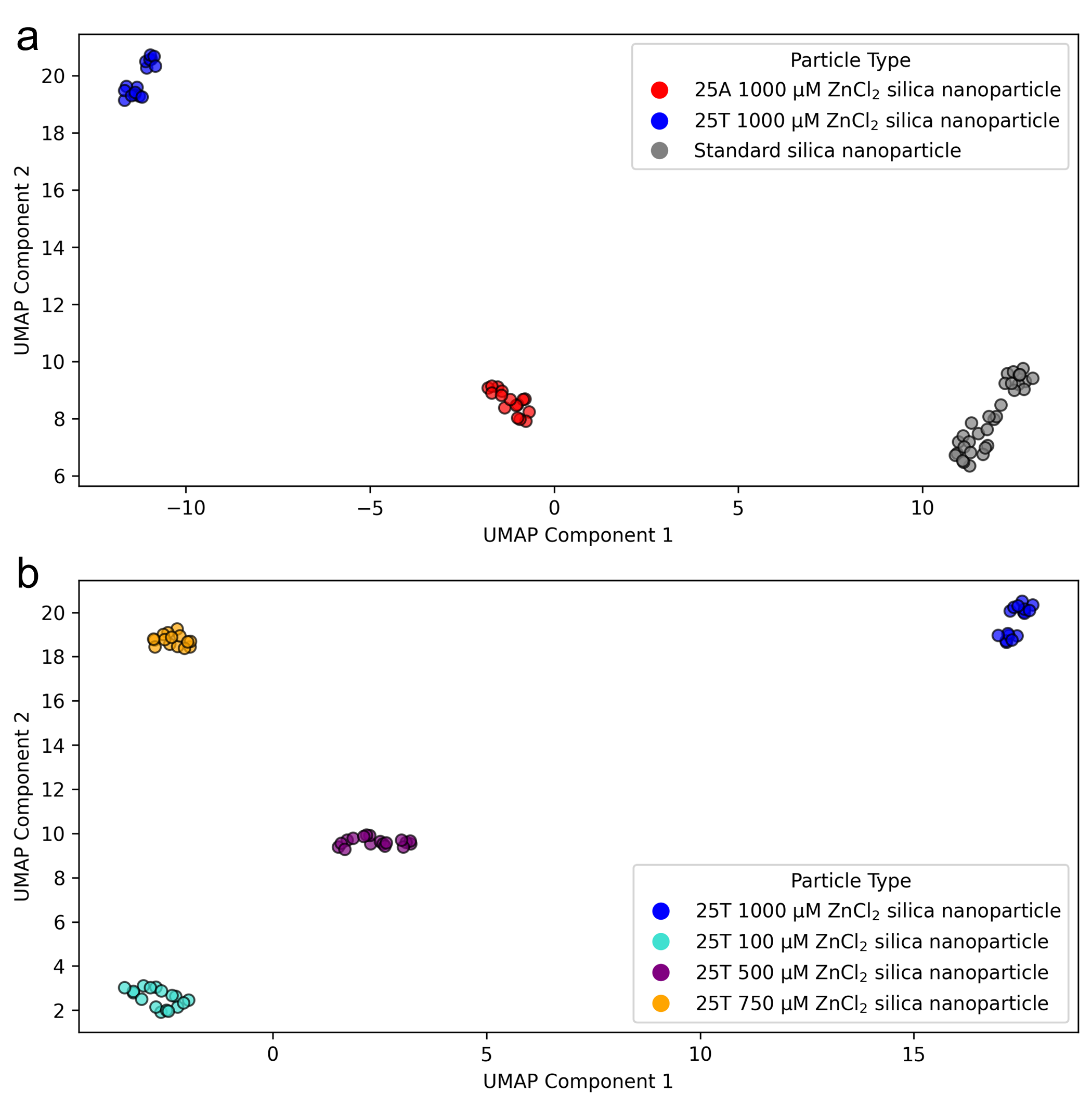}
    \caption{\textbf{2-Dimensional UMAP analysis of optically trapped silica nanoparticles.} (a) This UMAP scatter plot depicts the differences between the trapped 25A silica nanoparticles and 25T silica nanoparticles plotted alongside the standard silica nanoparticles. Both 25A silica nanoparticles and 25T silica nanoparticles use a 1000 $\mu$M concentration of ZnCl$_2$. (b) This UMAP scatter plot demonstrates the similarities and differences between varying the concentrations of ZnCl$_2$ on the binding of 25T to the silica nanoparticle surface. Both (a) and (b) use the following parameters of number of nearest neighbors = 50 and minimum distance = 0.0 to observe the global difference between particle types.}\label{fig3}
\end{figure*}

The raw PSD data were compared for each silica nanoparticle type. One candidate was selected from each group, and the f$_1$, f$_2$ and f$_3$ frequency peaks were displayed in sections for comparison. Both panels in Figure \ref{fig2} highlight differences observed between the particles. In Figure \ref{fig2}a, there is distinct separation in the PSD regarding width, amplitude and frequency of all peaks. Figure \ref{fig2}b shows differences between most frequency peaks for the 25T functionalized silica nanoparticles at different ZnCl$_2$ concentrations, the exception being the similarity between the 25T silica nanoparticles at 500 $\mu$M and 750 $\mu$M ZnCl$_2$ concentrations. These PSDs are very similar in frequency, amplitude and width at the f$_2$ and f$_3$ peaks.

A physics explanation for the observed trap frequency shift for the oligonucleotide base-coated silica nanoparticle in the levitated optomechanical trap is described in the Supporting Information. This model, which ignores the ZnCl$_2$ salt layer surrounding the surface of the silica nanoparticles, finds that the polarizability-to-mass ratio $(\alpha /m)$ changes depending on the oligonucleotide. It is this change in $\alpha /m$, not mass, which is responsible for the frequency shift. The measure of $\alpha /m$ is routinely used in metrology to complement mass spectrometry and sort fullerenes and polypeptides \cite{Gerlich2008,Ulbricht2008}. The estimates show that there is about a 1 kHz frequency shift per monolayer of DNA base molecule. There is a difference between adenine and thymine, and it is estimated that there is a frequency shift of 30 mHz corresponding to a single adenine molecule when compared to a single thymine base.

\begin{figure*}
    \centering
    \includegraphics[width=\linewidth]{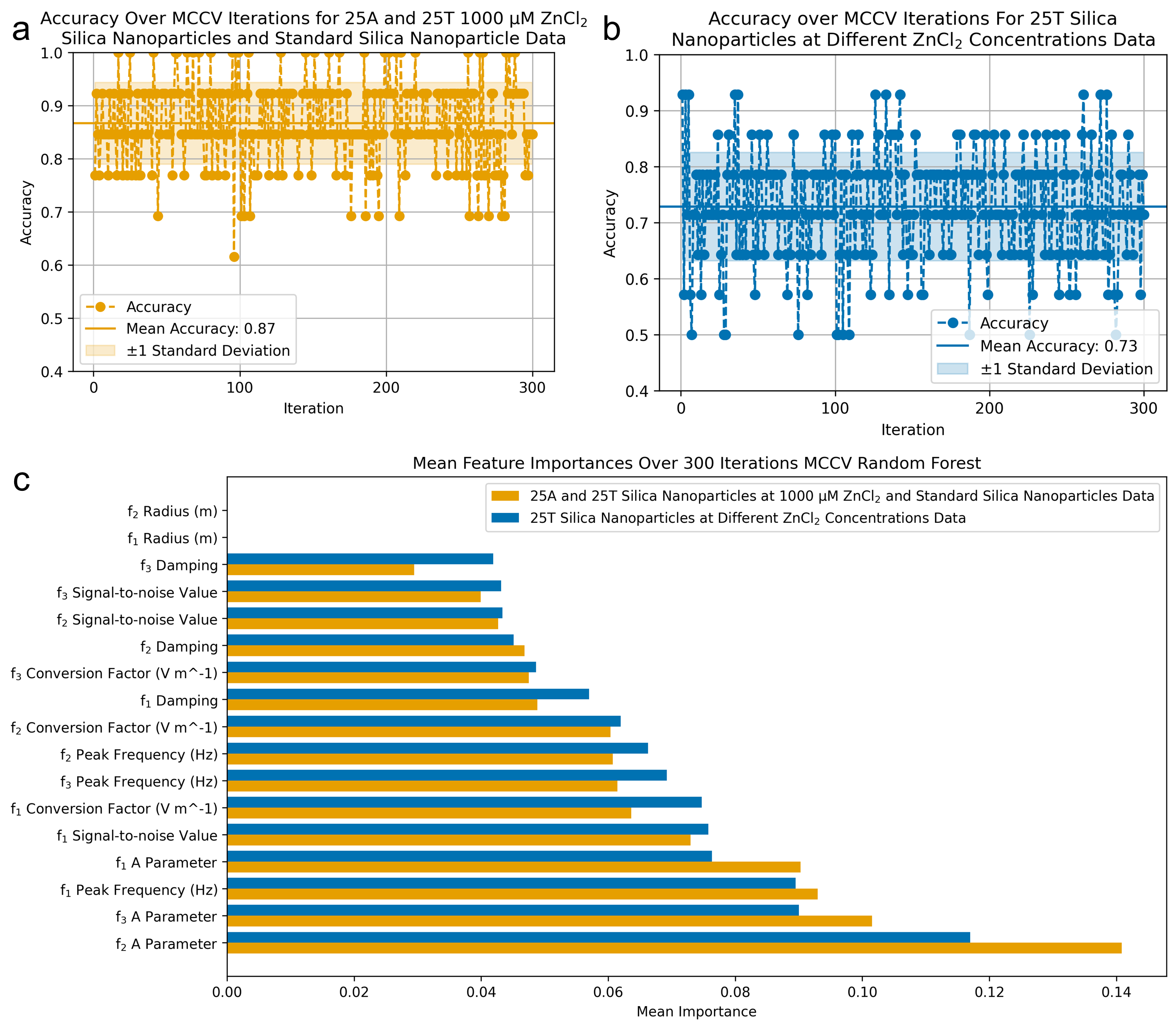}
    \caption{\textbf{Random forest results comparing two groups of silica nanoparticles.} (a) Random forest model accuracy for the 25A and 25T silica nanoparticles at 1000 $\mu$M ZnCl$_2$ and standard silica nanoparticles data over 300 MCCV iterations. (b) Random forest model accuracy for the 25T silica nanoparticles at different ZnCl$_2$ concentrations data over 300 MCCV iterations. (c) Mean feature importances of both datasets over 300 MCCV iterations.}\label{fig4}
\end{figure*}

Dimensionality reduction is often used in the field of machine learning to compress a dataset with many features, or columns, into a manageable number of components \cite{Jia2022}. This technique decreases the complexity of these data and improves the accuracy of classification \cite{Huang2019}. UMAP is a dimensionality reduction technique that can be run in a supervised learning mode to maximize the space between known classes in low-dimensional space that have features which are non-linearly correlated \cite{Allaoui2020,Pal2020}. This method was selected for its flexibility in the analysis of any type of high-dimensional data \cite{Becht2019}. Applying UMAP to the data collected allows examination of the differences between groups in a two-dimensional graphical representation.

The random forest technique is an ensemble machine learning algorithm that is effective as a generalized classification and regression model \cite{Biau2016}. Ensemble techniques have a greater accuracy than other machine learning methods, such as Support Vector Machines and K-Nearest Neighbors, because a group of classifiers tends to perform more accurately than an individual \cite{Rodriguez-Galiano2012,Sun2024}. A random forest is a group of decision trees, where each tree provides its own classification, and these are considered collectively through a vote to reach a classification consensus. The overall random forest algorithm considered the classification with the greatest number of votes from all the trees in the forest \cite{Jaiswal2017}. Usefully, the random forest can return a measure of feature importance \cite{Boulesteix2012}. Furthermore, the random forest model also produces an accuracy score, giving an indication of its performance. This is calculated using the Out of Bag Error which assesses the mean misclassification ratio of samples not used for training \cite{Petkovic2018}. 

To assess the performance of the random forest classifier in this experiment, a cross-validation technique was used. Accuracy scores from a single random forest run, particularly on a dataset of small sample size, are challenging to interpret and often do not give a complete picture. Monte Carlo cross-validation (MCCV) is a method which is suitable for small sample sizes. It functions by randomizing the samples in each training and test dataset and can be run for as many iterations as desired for robustness \cite{Xu2001}. MCCV can be applied to a random forest classification problem to validate the accuracy of the model, or optimize the features for use in future data \cite{Li2016}. A recent study compared resampling methods and found that no technique was consistently better than the others \cite{Nakatsu2023}.

The PSD comparison of one particle from each class in Figure \ref{fig2} suggests that there is a difference in optical properties between the silica nanoparticle groups. UMAP dimensionality reduction was used to plot the features of the particles from the Optoanalysis \cite{Rademacher2017} fitting in two dimensions.  The key objective of this analysis was to determine if it is possible to detect a difference between standard and oligonucleotide-functionalized silica nanoparticle. The clustering depicted in Figure \ref{fig3}a for 25A and 25T silica nanoparticles at the same 1000 $\mu$M concentration of ZnCl$_2$ suggests there is variation in the data that can be explained by silica nanoparticle status. The dimensionality reduction has considered all available features generated by the Lorentzian curve fitting to the raw data. There is a separation between the three particle types, with no overlap between these groups. Next, the UMAP method was applied to investigate if changing the concentration of ZnCl$_2$ has a detectable effect on 25T functionalization to the surface of silica nanoparticles. The separation observed in Figure \ref{fig3}b between all concentrations of ZnCl$_2$ shows agreement with data from the literature \cite{Huang2022} in that different concentrations of the binding agent can cause measurable changes in the quantity of DNA on the surface of silica nanoparticles. Both 2-Dimensional UMAP clustering figures demonstrate the effectiveness of this method in clustering individual groups and separating the particle type classes.

A random forest model was trained on the Optoanalysis tool parameters with outliers removed. The datasets were split into 80\% for training and 20\% for testing. Training parameters and features were iterated through, and the optimal combination was selected to the deliver the best model accuracy. Due to the limitation in size of the datasets, where the 25A and 25T silica nanoparticles at 1000 $\mu$M ZnCl$_2$ and standard silica nanoparticles data has 64 entries, and the 25T silica nanoparticles at different ZnCl$_2$ concentrations data have 66 entries, the MCCV method was utilized to give a more complete understanding of random forest performance. Figure \ref{fig4}a shows that the random forest model performs well over the 300 iterations, at best there is perfect accuracy, at worst it is 62\% accurate, the mean accuracy is 87\%. This shows agreement with the UMAP plot in Figure 3a with distinct clustering between particle types suggesting  that classification not random. In Figure \ref{fig4}b the model has a good mean accuracy at 73\%, with a minimum of 50\% and maximum of 93\%. An explanation for the weaker mean accuracy in Figure \ref{fig4}b compared to Figure \ref{fig4}a could be the similarity between the DNA binding for 25T 500 $\mu$M and 750 $\mu$M ZnCl$_2$ silica nanoparticles \cite{Huang2022}. The ranking of mean feature importance in Figure \ref{fig4}c is in the same order for both models, however the magnitude of importance varies. The f$_2$ and f$_3$ A parameters are most important, with the f$_1$ and f$_2$ radii coming in last. This suggests a consistent importance of these features to classify the particle groups. The f$_3$ radius was not used in any iteration and is therefore not shown.

Transmission electron microscopy (TEM) visualization of DNA molecules is challenging\cite{Kabiri2019}. One study was able visualize DNA duplex features such as the major groove, minor groove and helix pitch using high-resolution TEM at 70 keV, however further information about base sequence was challenging to infer \cite{Marini2015}. There are heavy metal staining methods to visualize DNA with TEM including uranyl acetate, but this is difficult due to its radioactivity \cite{DeCarlo2011}. TEM imaging of silica nanoparticles is straightforward as the particles provide good contrast for visualization \cite{Ibrahim2010}.

\begin{figure}
    \centering
    \includegraphics[width=\linewidth]{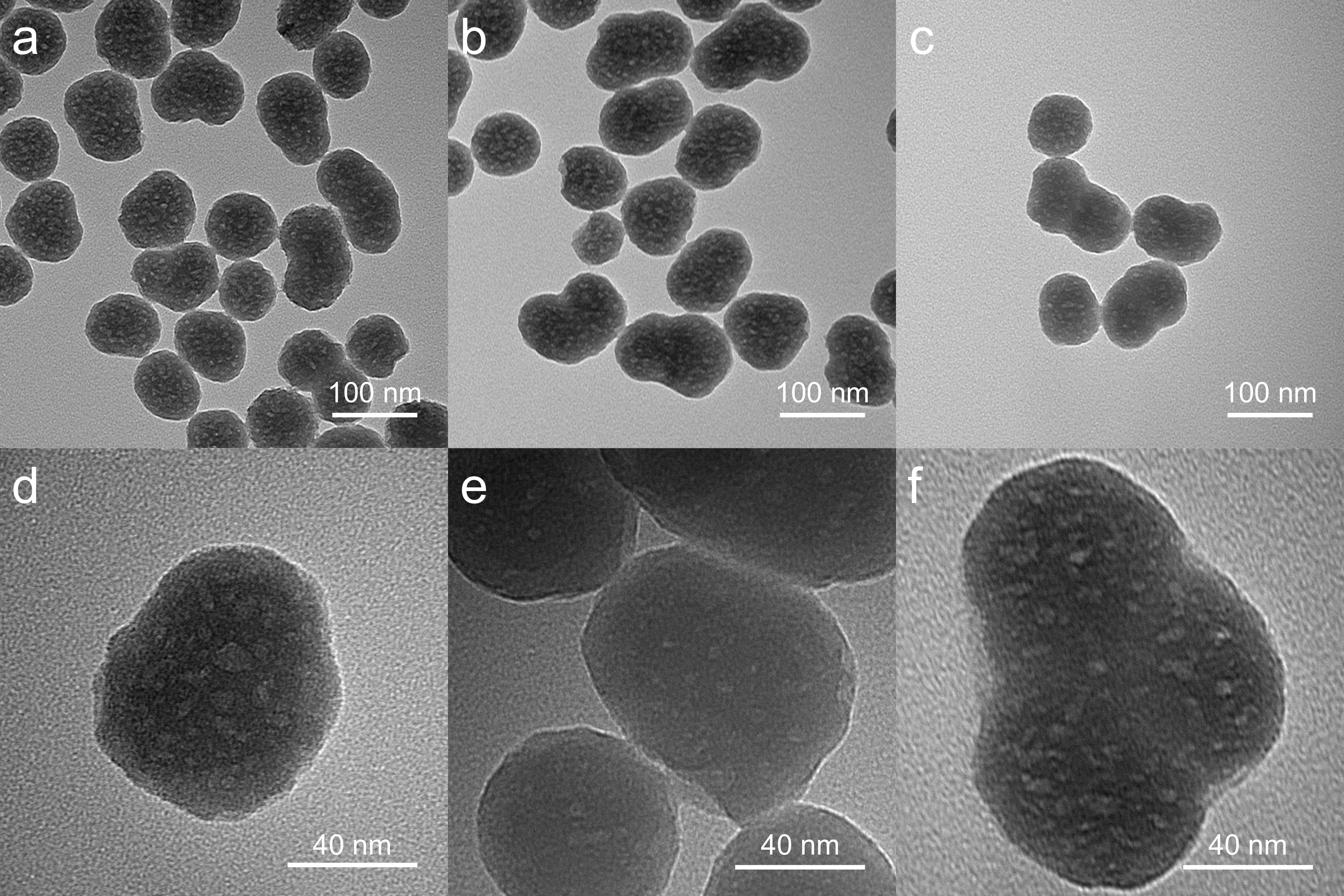}
    \caption{\textbf{TEM images of silica nanoparticles.} (a) 25T 100 $\mu$M ZnCl$_2$ silica nanoparticles. (b) 25T 500 $\mu$M ZnCl$_2$ silica nanoparticles. (c) 25T 1000 $\mu$M ZnCl$_2$ silica nanoparticles. (d) Standard silica nanoparticle. (e) 25A 1000 $\mu$M ZnCl$_2$ silica nanoparticles. (f) 25T 750 $\mu$M ZnCl$_2$ silica nanoparticle. a – c are at 100,000 $\times$ magnification. d – f are at 600,000 $\times$ magnification.}\label{fig5}
\end{figure}

All particle types that were tested in the optical trap were also imaged using TEM. The same sample preparation method was used as shown in Figure \ref{fig1}. The images presented in Figure \ref{fig5} show the variance in particle size and shape. This is the reason why outliers were removed in the data analysis since considerable differences could be due to physical rather than optical properties. Despite sonicating the particle solutions, there are still clusters of particles present during imaging. The speckled white surface texture is representative of the surface roughness of the particles; they are not perfect spheres. There were no clear visual differences between the nanoparticle groups.

The analysis performed in this study found detectable differences in optical properties between the types of silica nanoparticles. Taking one example from each particle group and plotting the raw data demonstrates that the Lorentzian curve fitting to produce the UMAP plots and random forest models is based upon foundational differences in their inputs. The UMAP result indicates that there are differences in the features between the silica nanoparticle classes that cause observable separation. The random forest modeling and MCCV iterations reveal that these classes can be identified with high mean accuracies of 0.87 and 0.73. Finally, the TEM imaging suggests that there is not an observable difference between the groups of silica nanoparticles.

To overcome the indistinguishable differences between silica nanoparticle groups using TEM, future work could include the use of Scanning Electron Microscopy and Energy Dispersive X-ray Spectroscopy (SEM-EDS). This technique can detect the presence of elements in a sample of atomic number greater than eleven\cite{DeTata2023}, so would detect the addition of zinc, carbon and phosphorus on the DNA-functionalized silica nanoparticles when compared to standard silica nanoparticles.

There is a limitation in data quantity as can be seen in the UMAP plots in Figure \ref{fig3}. To extend this analysis, a larger dataset would need to be generated. However, the MCCV result regarding iterating through different variations of training and testing data does give confidence that there is a real difference in the data that are classified by the random forest model.

Having a classifier to distinguish between DNA strands could have applications in diagnostic testing, particularly where current sequencing technologies are challenged by GC-rich or repetitive regions \cite{Browne2020,Morton2020}. 25A and 25T oligonucleotides were selected in this study because they have been shown to bind well with Zn$^{2+}$ to silica nanoparticles \cite{Huang2022}, however, they are also as different from each other as possible. Changing the bases to 24T with one A, swapping around the position of the A nucleotide in the strand, or changing the strand length and sequence entirely would be necessary to demonstrate this approach for further applications. Another question not investigated in this study was the effect of duplex DNA. Experimenting with a 25A-25T duplex would also indicate if there was a uniqueness to the DNA structure. Perhaps the additional binding offered by another phosphodiester backbone or the greater mass would affect how the data is clustered.

In summary, this study could lead down several directions. The frequency shift of 30 mHz in the proposed model can be feasibly detected in future experiments and will need to be investigated. If this is proved experimentally, then the ability to detect individual mass changes due to differences in base sequence could have uses in determining mutational or methylation modifications to DNA strands \cite{Edwards2005,Ehrich2005}. Further development requires the correct foresight into how this technology could be applied. Work and thought are needed to explore why these results are present. Exploration into the underlying cause of different optical properties may shed light on the reasoning behind the UMAP clustering and random forest classification. Bioinformatics is becoming an ever more essential component of modern medicine; perhaps this optical trap classifier will be another method in the toolkit.

\subsection*{Acknowledgments} 
We express gratitude to Laura Barbara and Jack Homans for their support in configuring the optical trap. The authors thank Eugen Stulz for his assistance in the preparation of the functionalized particles and Ahmed Dawoud for his guidance on the random forest analysis. The authors are grateful to Regan Doherty for the training on the TEM.

We acknowledge funding from the EU Horizon Europe EIC Pathfinder project QuCoM (10032223), from the UK funding agency EPSRC (grants EP/W007444/1, EP/V035975/1, EP/V000624/1), and from the Leverhulme Trust (RPG-2022-57).



\clearpage

\section*{Supplementary Information}


\subsection*{Particle preparation}

Deionized water was used in all experiments for dilution to ensure that no potential contaminants such as DNases would break down the oligonucleotides. All particle solutions were prepared using 100 $\mu$L of 100 nm diameter Corpuscular Silicon Dioxide Nanospheres at 5.0\% concentration diluted to a final volume of 1100 $\mu$L. Following the addition of silica nanoparticles, Amcor M Parafilm was wrapped around the vial to create an airtight seal, slowing down the rate of particle aggregation. Volumes were measured using Eppendorf Research Plus micro-pipettes with tips replaced after each quantity of solution was extracted. Silica nanoparticles of volume 100 $\mu$L were pipetted into 1.5 ml microcentrifuge tubes and placed in a James Products Europe Sonic 3MX Professional Ultrasonic Cleaner sonicator at 37 kHz for 30 minutes at the lowest temperature setting of 17 °C to further reduce the aggregation of silica nanoparticles. Otherwise, there could be observed rotational effects in the spectrum which are undesired. After sonication, the other solution components were added in the following order: water, DNA (if used) and ZnCl$_2$ (if used). The tubes were inverted ten times after each solution was added to ensure thorough mixing. To keep the 25A and 25T oligonucleotides intact, only the silica nanoparticles were placed in the sonicator.

A stepwise approach was taken to account for increasing the complexity of functionalizing the surface of silica nanoparticles. The first step was to trap silica nanoparticles without any surface modifications. This acted as a control and was carried out to ensure proper functionality and alignment of the optical setup. In the second phase, a 100 mM ZnCl$_2$ stock solution was prepared by dissolving 1.363 g of Sigma Aldrich $\geq$ 98\% reagent grade ZnCl$_2$ in 100 mL deionized water, then ZnCl$_2$ was added to the silica nanoparticles at a final concentration of 1000 $\mu$M. The third stage involved addition of 25A (a 25-mer of deoxyadenosine monophosphate) or 25T (a 25-mer of deoxythymidine monophosphate) to the silica nanoparticle solution with ZnCl$_2$ and then optically trapped. These were selected due to the ability to functionalize silica nanoparticles with fluorescently tagged 25As \cite{Huang2022}. 25T was also chosen as it is vastly different in structure to 25A and also proved successful in functionalizing to silica nanoparticles \cite{Huang2022}. Both oligonucleotides were ordered from Integrated DNA Technologies (IDT) and resuspended in deionized water to make a 10 $\mu$M stock solution. The molecular weight of 25A is 7,768.3 g mol$^{-1}$ and was measured by IDT as having an optical density at the 260 nm wavelength (OD$_{260}$) of 169.7 which is equal to 559.3 nmol. 559 $\mu$L of deiozined water was used to resuspend the 25A to make a 10 $\mu$M stock solution. 25T has a molecular weight of 7,542.9 g mol$^{-1}$. IDT measured the OD$_{260}$ as 117.5 equal to 578.3 nmol. Here, 578 $\mu$L of deionized water was pipetted to resuspended the 25T to create the 10 $\mu$M stock solution. As the concentration of oligonucleotides at 400 nM and ZnCl$_2$ at 1000 $\mu$M yields the maximum loading capacity onto the silica nanoparticles \cite{Huang2022}, this research used a 400 nM concentration of oligonucleotides in all samples. This relatively low concentration reduced the likelihood of the nebulizer mesh piece becoming clogged. A final step involved combining 25T silica nanoparticles with the following additional concentrations of ZnCl$_2$: 100, 500 and 750 $\mu$M. At all stages in the experiment, the oligonucleotides were refrigerated to reduce the rate of degradation.

\begin{figure}
    \centering
    \includegraphics[height=80mm]{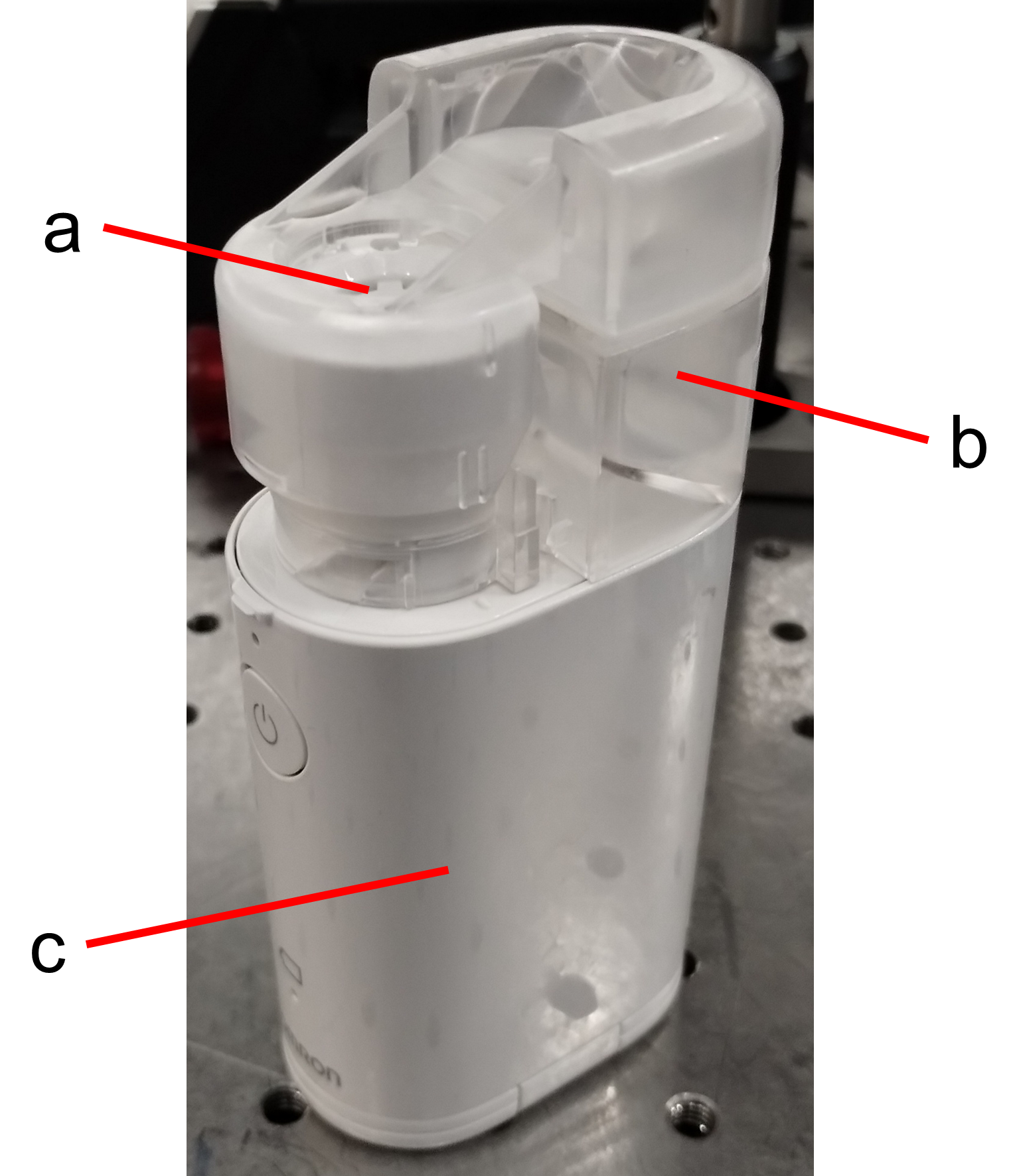}
    \caption{Supplementary Figure 1: \textbf{The Omron MicroAIR U100 nebulizer.} There are three main parts: (a) is the interchangeable mesh piece where the aerosol droplets are released, (b) is the reservoir where nanoparticle solutions are placed, and (c) is the battery power source compartment.}\label{figA1}
\end{figure}

An Omron MicroAIR U100 Portable Nebulizer (Supplementary Figure~\ref{figA1}) was used to release particles into the optical trap. Following each experiment, the reservoir and head mesh pieces were rinsed using deionized water to prevent silica nanoparticles clogging up the mesh such that a sufficient stream of aerosol would be released. Between changing over solution types, the mesh piece was replaced to avoid contamination, particularly of importance when considering the 25A and 25T functionalized particle stages. A particulate respirator was worn at all stages when the nebulizer was in use to avoid inhalation of silica nanoparticles.

\subsection{Optical trap laser setup}

\begin{figure}
    \centering
    \includegraphics[width=\linewidth]{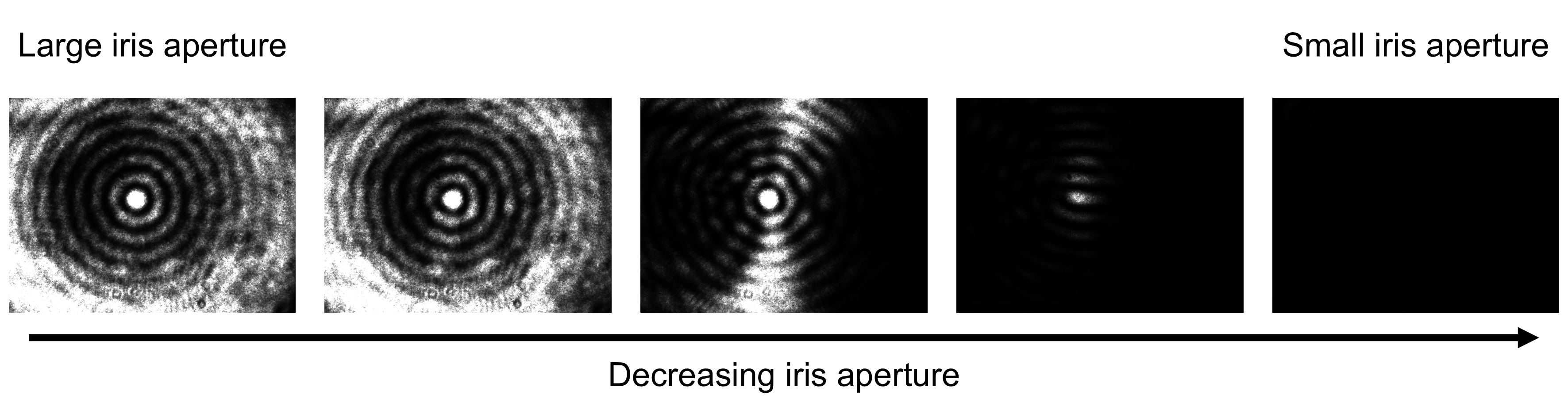}
    \caption{Supplementary Figure 2: \textbf{IR images of decreasing iris apertures:} A series of images demonstrating the removal of the Poisson spot with decreasing iris aperture. Closing the iris aperture decreases the size of the Poisson spot until the aperture size is equal to the diameter of the parabolic mirror.}\label{figA2}
\end{figure}

Previous optical trap experiments \cite{Vovrosh2018} used a setup similar to the one in this work. The iris was included such that when the aperture is reduced, the high-power Poisson spot is removed (Supplementary Figure~\ref{figA2}), decreasing the signal-to-noise ratio which increases the position resolution of the trapped particle that can be detected by the photodiode. This is necessary as the bright spot in the center of the beam overlaps the scattered field (E$_{scat}$) and diverging field (E$_{div}$). The bright Poisson spot forms from the flat mirror edges around the parabolic mirror and was used to confirm the laser alignment. The images in Supplementary Figure~\ref{figA2} were captured by a Point Gray Research CMLN-13S2M-CS infrared camera.

The optical setup is shown in Supplementary Figure~\ref{figA3}. Light comes out of a 40 mW NKT Photonics Koheras Basik Mikro 1550 nm laser and is seeded into a Nuphoton High Power Erbium-Doped Fibre Amplifier (EDFA). This is then released from a Thorlabs 3.0 mm diameter collimator and reflects off a mirror, then travels through a Thorlabs $\lambda$/2 @ 1550 nm MULTI-ORDER waveplate set at an angle of 38° to control the power of the trapping laser beam. The polarizing beam splitter (PBS) trapping beam then enters the Thorlabs $\lambda$/4 @ 1550 nm MULTI-ORDER waveplate set at an angle of 58° to control the polarization of the light and the direction of the diverging and scattered light (E$_{div}$ + E$_{scat}$). The trapping beam then reflects off an alignment mirror and passes through an iris. Upon the trapping beam entering the vacuum chamber, the light is reflected off a parabolic mirror. This is necessary to tightly focus the beam onto the trapped particle to achieve a higher laser intensity gradient. The E$_{div}$ + E$_{scat}$ beam then travels back through the iris, alignment mirror and $\lambda$/4 waveplate and is sent at right angles to the trapping beam as it goes through the PBS. Following reflection from the final alignment mirror, the E$_{div}$ + E$_{scat}$ beam enters a Thorlabs NDC-50C-4M variable ND filter. This prevents saturation of the Thorlabs 800 – 1700 nm DC – 5 MHz photodetector. The signal is finally displayed on the Rohde and Schwarz RTO2014 oscilloscope.

\begin{figure}[htbp]
    \centering
    \includegraphics[width=\linewidth]{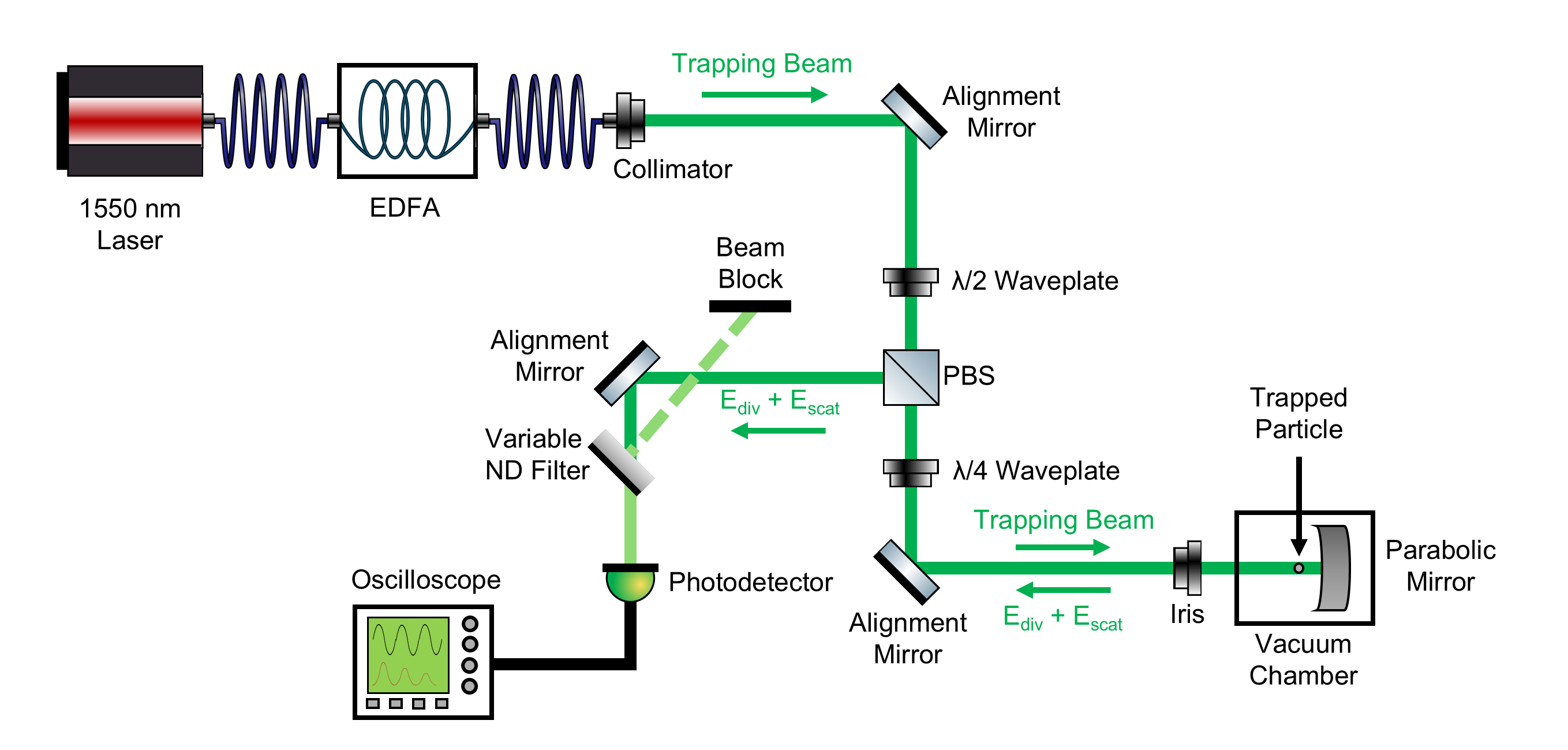}
    \caption{Supplementary Figure 3: \textbf{Optical setup:} A diagrammatic representation of the optical setup with the laser system and optical beam path, focusing parabola in vacuum chamber and data collection by an oscilloscope.}\label{figA3}
\end{figure}

The laser’s power was measured between the alignment mirror and iris using a ThorLabs S146C Photodiode Power Sensor coupled to a ThorLabs PM100D Handheld Optical Power and Energy Meter Console. The power was measured over 100 samples with a mean value of 395.4 mW, a standard deviation of 2.537 mW, a minimum value of 391.3 mW and a maximum value of 400.7 mW.

After particles were released into the vacuum chamber, the pressure was reduced using a vacuum pump. Pressure was measured using an Agilent Technologies FRG-720 pressure gauge. For each experiment, the raw oscilloscope data was aimed to be saved at 3.5 mbar as the PSD produced clean peaks on the oscilloscope for the Lorentzian curve fitting at this pressure. However, due to the manual operation of the controls, it was not possible for the pressure to be exactly 3.5 mbar each time. However, as the pressure gauge has an accuracy of ± 15\%, the pressures at which data was saved was within the range of error.

\subsection{Data collection and analysis}

Data are freely available from the University of Southampton Institutional Repository \cite{datarepository}. Following the collection of raw data from the oscilloscope, the frequency plot of the time domain signal was obtained using a Fourier transform. This was carried out in Python code using the Optoanalysis package \cite{Rademacher2017}. A Graphical User Interface (GUI) was developed on top of this code to enable seamless data entry as seen in Supplementary Figure~\ref{figA4}a. Following the display of the PSD, the recorded pressure was entered, as was the estimated peak frequency of the z, x and y degrees of motion, as well as a Yes/No value if the PSD Lorentzian curve was a good fit (Supplementary Figures~\ref{figA4}b-d). The data for each trapped particle was recorded as a row in a comma-separated value (CSV) table.

\subsection*{Silica nanoparticles analyzed}
\begin{table}[htbp]
\centering
\caption{\textbf{Supplementary Table 1}. Values for the particle type and number analyzed in the data analysis section.}
\begin{tabular}{c|c}
Particle Type & Number Analyzed \\
\hline
Standard silica nanoparticle & 32 \\
25A 1000 $\mu$M ZnCl$_2$ silica nanoparticle & 16 \\
25T 1000 $\mu$M ZnCl$_2$ silica nanoparticle & 16 \\
25T 100 $\mu$M ZnCl$_2$ silica nanoparticle & 17 \\
25T 500 $\mu$M ZnCl$_2$ silica nanoparticle & 18 \\
25T 750 $\mu$M ZnCl$_2$ silica nanoparticle & 15 \\
\end{tabular}
\label{tableS1}
\end{table}

\begin{figure}[h!]
\includegraphics[width=\linewidth]{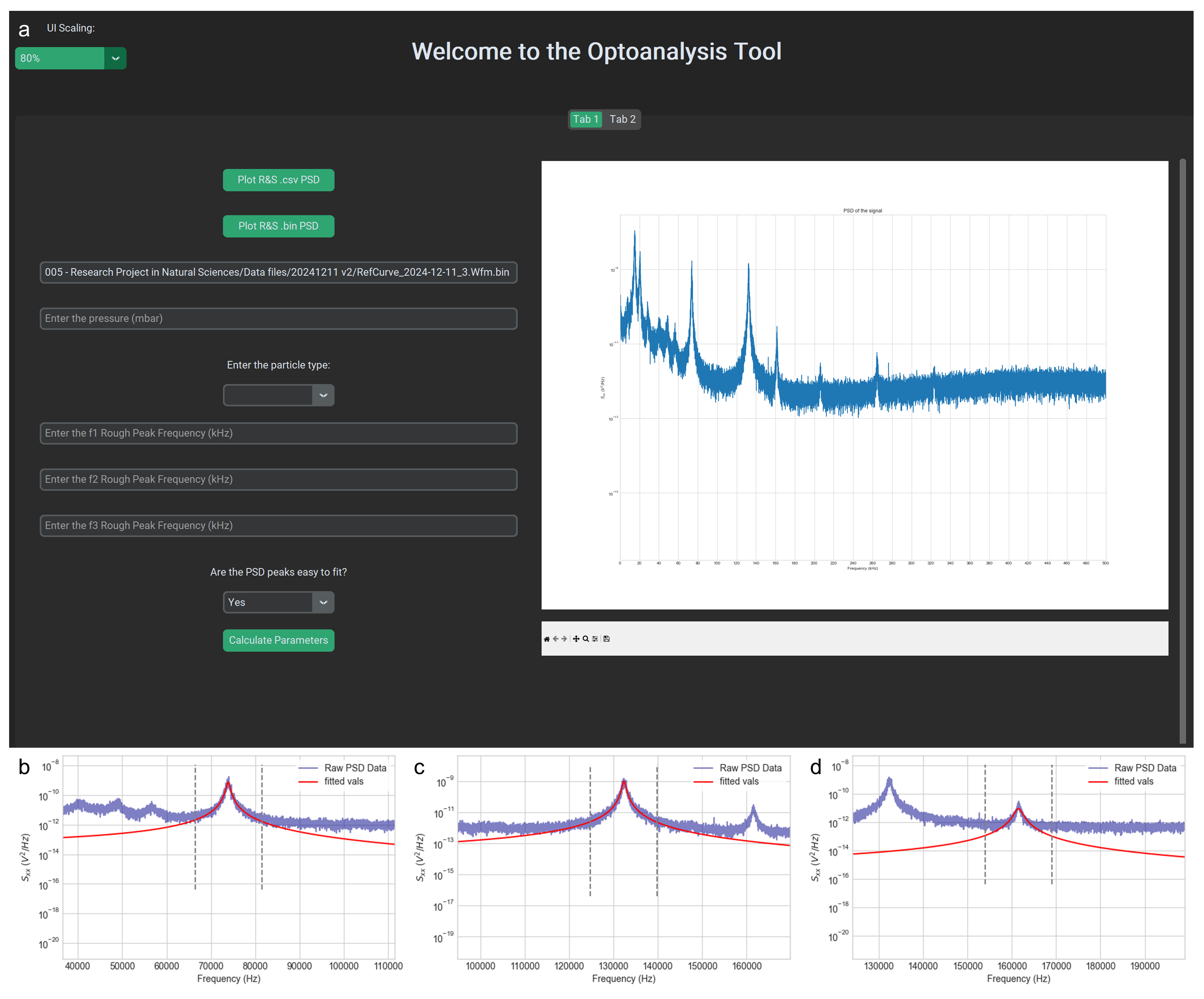}
\centering
\caption{Supplementary Figure 4: \textbf{GUI of Optoanalysis} with an example PSD and Lorentzian curve fitting. (a) Screenshot of the Optoanalysis Tool GUI Developed to Enable Data Entry into a CSV Table, (b) Lorentzian curve fitting to the f$_1$ peak, (c) Lorentzian curve fitting to the f$_2$ peak and (d) Lorentzian curve fitting to the f$_3$ peak.}\label{figA4}
\end{figure}


\subsection{Process of TEM imaging}
The TEM imaging samples were prepared using the same method as for the optical trapping previously described. 5 $\mu$L of each sample was pipetted onto a copper grid and left to set for three minutes. Then, the excess solution was wicked away with filter paper. The copper grid with the sample was loaded into a holding rod and inserted into the TEM for imaging. All images were taken using a Hitachi HT7800 TEM.

\begin{table*}
    \centering
    \begin{tabular}{c|c|c|c}
Nucleobase & Polarizability $\alpha$ (C·m$^2$/V) & Mass m (kg) &  $\frac{\alpha}{m}$ (C·m$^2$/V·kg)\\
\hline
Adenine    & $1.65 \times 10^{-39}$               & $2.24 \times 10^{-25}$ &  7.73 $\times 10^{-15}$ \\                                        
Thymine    & $1.48 \times 10^{-39}$               & $2.09 \times 10^{-25}$ &  7.08 $\times 10^{-15}$ \\                                         
    \end{tabular}
    \caption{\textbf{Literature values} for mass and polarizability for adenine and thymine from density functional theory (DFT) and Raman scattering experiments.}
    \label{tableS1}
\end{table*}

\subsection*{Oscillation frequencies of polarizable particle in optical dipole trap:} This part derives a physics reason for the observed trap frequency shift for the functionalized SiNPs. The reason is a small change in the polarizability to mass ratio $\left( \frac{\alpha}{m}\right)$ for coated SiNPs. 

{\bf Trap frequency formula for \texorpdfstring{$z$}{z}-direction:}
For a diffraction-limited focus of a Gaussian laser beam which is forming the optical trap, the trap frequency along the ($z$) trapping direction is given according to the ponderomotive light-matter interaction by:
\begin{equation}
    \omega_{\text{trap}} = \sqrt{\left( \frac{\alpha}{m}\right) \frac{4 P}{\pi \epsilon_0 c w_0^4}} ,
\end{equation}
with  $P$ the incident trapping laser power, $w_0$ the beam waist radius at the focal point, $m$ is the mass of the nanoparticle, $\alpha$ its static polarizability, $\epsilon_0$ dielectric constant of vacuum, and $c$ is the speed of light.

For a laser power $P$ = 0.1~W, a beam waist of $w_0$ = 1 $\mu$m = $1 \times 10^{-6}$m and for adenine and thymine mass and static polarizability (e.g. Table~\ref{tableS1}), this leads to a $z$-trap frequency $f_0 = 135\, \text{kHz}$, in good agreement with the experimentally observed values.

{\bf Dielectric properties of oligonucleotides} Typically, the static (frequency-independent) polarizabilities can be computed as response properties or finite field calculations. There have been extensive ab-initio density functional theory (DFT) calculations using various sets of density functionals \cite{hafner2008ab, yang2019quantum, zhao2022density} and complemented by Hartee-Fock methods \cite{gowtham2007physisorption}, and classical electrodynamics calculations by Kramers-Kronig relations \cite{pinchuk2004optical, fichou2008handbook}. 

Theoretical values for static polarizabilities can then be compared to experiments on dielectric properties of oligos for instance by Rayleigh scattering \cite{jonin2021hyper}, by electrostatic force microscopy \cite{cuervo2014direct}, or by optical extinction coefficient measurements \cite{banihashemian2013optical}. Table~\ref{tableS1} shows the values for $\alpha$ and m for adenine and thymine used in this work. 

{\bf Silica nanoparticle (SiNP) parameters used:} the silica (SiO$_2$) particle diameter $d = 100$~nm results in a particle mass of $m_{silica} = 1.15 \times 10^{-18}$~kg and a static polarizabilty as estimated using Clausius-Mossotti relation for a silica refractive index at 1550 nm to be $n \approx 1.44$. This results in a static polarizability of $\alpha_{silica} = 3.66 \times 10^{-33}$ C $\cdot$ m$^2$/V (SI units).

\subsection{Trap frequency change for oligo coated SiNPs}
One monolayer of adenine gives about 1~kHz of frequency shift (0.74\% shift in trap frequency with respect to the uncoated silica case) in the optical trap of a 100 nm diameter silica nanoparticle. The increase in monolayers (ML) is linear in the frequency shift in the regime of small perturbations. These are small effects, but resolvable by levitated optomechanics. The calculation goes as follows.

{\bf Parameters for 1 monolayer (1ML) of adenine on a SiNP:} The number of adenine molecules in monolayer coating of a spherical 100~nm diameter silica nanoparticle is $N = 3.14 \times 10^{4}$. The total polarizability of 1ML of adenine is:
$\alpha_{A,total} = N \times \alpha_{A} = 5.18 \times 10^{-35} \, C \cdot m^2 / V$. The mass of one Adenine monolayer:
$m_{A} = N \times 2.24 \times 10^{-25}$~kg $ = 7.03 \times 10^{-21} \,$~kg.
Total polarizability of an adenine coated nanoparticle is:
$\alpha_{total} = \alpha_{silica} + \alpha_{A,total} = 3.66 \times 10^{-33} + 5.18 \times 10^{-35} \approx 3.71 \times 10^{-33}.$ We derive a corresponding trap frequency ratio of $\frac{\omega_{total}}{\omega_{silica}} = 1.007$. By analogous calculation, we derive a smaller frequency shift for 1ML of thymine, namely $\omega_{T}=0.93\omega_{A}$. This means adenine and thymine can be distinguished.

\subsection{Frequency shift for a single adenine/thymine molecule}
We now estimate the limiting case of the expected frequency shift for the case of a single oligonucleotide molecule adsorbed to the SiNP surface. The frequency shift for the limit of small perturbations is derived by the trap frequency ratio of with the molecule adsorbed $\omega$ versus without the molecule adsorbed $\omega_0$ is given by:
$$
\frac{\omega}{\omega_0} = \sqrt{\frac{\alpha_{silica} + \alpha_{molecule}}{m_{silica} + m_{molecule}} \bigg/ \frac{\alpha_{silica}}{m_{silica}}} .
$$
Since $m_{molecule} \ll m_{silica}$, we approximate:
$$
\frac{\omega}{\omega_0} \approx \sqrt{1 + \frac{\alpha_{molecule}}{\alpha_{silica}}} ,
$$
and the relative frequency change:
$$
\frac{\Delta \omega}{\omega_0} \approx \frac{1}{2} \frac{\alpha_{molecule}}{\alpha_{silica}}.
$$
This results in relative frequency shifts for a single adenine molecule of:
$\frac{\Delta \omega}{\omega_0} \approx 2.25 \times 10^{-7}$, which corresponds to an absolute frequency shift of, $\Delta f_{A} = f_0 \frac{\Delta \omega}{\omega_0} = 0.030$~Hz, and for thymine to $\
\Delta f_{T} = 0.028$ Hz. We expect this shift to be resolvable in future optimized optical levitation setups with stabilized power spectral density (PSD) features.

\bibliography{main}

\end{document}